\PassOptionsToPackage{table,xcdraw}{xcolor}
\documentclass[10pt,conference]{IEEEtran}
\IEEEoverridecommandlockouts

\usepackage{multirow}
\usepackage[most]{tcolorbox}
\usepackage{pifont}
\usepackage{array} 
\usepackage{helvet} 
\usepackage{hyperref}
\usepackage[disable]{todonotes}
\usepackage{fancyvrb}
\usepackage{xurl}
\usepackage[shortlabels]{enumitem}
\usepackage{soul} 
\usepackage{graphicx}
\usepackage{svg}
\usepackage{mdframed}
\usepackage{cite}
\usepackage{amsmath,amssymb,amsfonts}
\usepackage{algorithmic}
\usepackage{textcomp}
\def\BibTeX{{\rm B\kern-.05em{\sc i\kern-.025em b}\kern-.08em
    T\kern-.1667em\lower.7ex\hbox{E}\kern-.125emX}}

\usepackage{flushend}
\usepackage{balance}
\usepackage{multicol}
\usepackage{colortbl}
\usepackage{graphicx}
\usepackage{amssymb}
\usepackage{booktabs}
\usepackage{tikz} 
\usepackage{geometry}
\definecolor{darkgreen}{RGB}{0,160,0}
\definecolor{gold}{rgb}{1.0, 0.84, 0.0}

\usepackage{listings}
\newcommand{\xmark}{\textcolor{red}{\ding{55}}}    

\usepackage[switch]{lineno}

\begin{document}




\title{Evaluating Software Development Agents: Patch Patterns, Code Quality, and Issue Complexity in Real-World GitHub Scenarios
}

\author{
    \IEEEauthorblockN{Zhi Chen, Lingxiao Jiang}
    \IEEEauthorblockA{
        Centre for Research on Intelligent Software Engineering\\
        School of Computing and Information Systems\\
        Singapore Management University\\
        \{zhi.chen.2023, lxjiang\}@smu.edu.sg
    }
    \vspace{-2.5em}
}

\maketitle

\begin{abstract}
In recent years, AI-based software engineering has progressed from pre-trained models to advanced agentic workflows, with Software Development Agents representing the next major leap. These agents, capable of reasoning, planning, and interacting with external environments, offer promising solutions to complex software engineering tasks. However, while much research has evaluated code generated by large language models (LLMs), comprehensive studies on agent-generated patches, particularly in real-world settings, are lacking. This study addresses that gap by evaluating 4,892 patches from 10 top-ranked agents on 500 real-world GitHub issues from SWE-Bench Verified, focusing on their impact on code quality. Our analysis shows no single agent dominated, with 170 issues unresolved, indicating room for improvement. Even for patches that passed unit tests and resolved issues, agents made different file and function modifications compared to the \textit{gold patches} from repository developers, revealing limitations in the benchmark’s test case coverage. Most agents maintained code reliability and security, avoiding new bugs or vulnerabilities; while some agents increased code complexity, many reduced code duplication and minimized code smells. Finally, agents performed better on simpler codebases, suggesting that breaking complex tasks into smaller sub-tasks could improve effectiveness. This study provides the first comprehensive evaluation of agent-generated patches on real-world GitHub issues, offering insights to advance AI-driven software development.

\end{abstract}


\section{Introduction}

\textit{\textbf{Background:}} \textit{Agents Are The Future Of AI}~\cite{toews2024agents}. AI-based software engineering has evolved rapidly, moving from pre-trained models~\cite{yu2024codereval} to fine-tuned large language models (LLMs)~\cite{fan2023automated}, in-context learning~\cite{geng2024large}, and further advancing with techniques like chain-of-thought~\cite{li2023structured} and agentic workflows~\cite{hong2024metagpt, qian2024chatdev}. \textit{Software Development Agents} represent the next step in AI development \cite{hong2024metagpt,qian2024chatdev, hou2023large}, integrating reasoning, planning, and interaction with external environments to perform autonomous tasks and make decisions which enables them to tackle complex software engineering challenges beyond simple function generation. Emerging agents, such as Amazon Q Developer
and EPAM AI/Run Developer Agent, highlight AI's potential to address more complex development tasks, marking a new direction where agentic workflows drive the creation of sophisticated, real-world applications.


\textit{\textbf{Motivation:}} While software development agents have advanced rapidly, comprehensive evaluations of the code they generate in real-world tasks are still lacking. Agents differ from large language models (LLMs) that generate code from static prompts by incorporating reasoning, planning, and interactions with external environments, which requires a distinct evaluation approach. Although many studies have evaluated LLM-generated code, focusing on aspects such as security vulnerabilities, memorization risks, and reliability~\cite{pearce2022asleep, liu2024no, asare2024user, siddiq2024quality, liu2024reliability, chen2024promise}, these findings may not directly apply to agents due to their more complex workflows and autonomous decision-making. Furthermore, much of the existing research is based on simpler tasks like generating Python functions or solving algorithmic challenges~\cite{liu2024refining,liu2024no, zhuo2024bigcodebench}, or on controlled vulnerability scenarios~\cite{pearce2022asleep,majdinasab2024assessing}, which do not capture the complexity of real-world software development. Our study addresses this gap by evaluating agent-generated patches on real GitHub issues, providing insights that are more relevant to real-world software development.

\vspace{-0.5em}
\begin{figure}[htbp]
    \centering
    \includegraphics[width=\columnwidth]{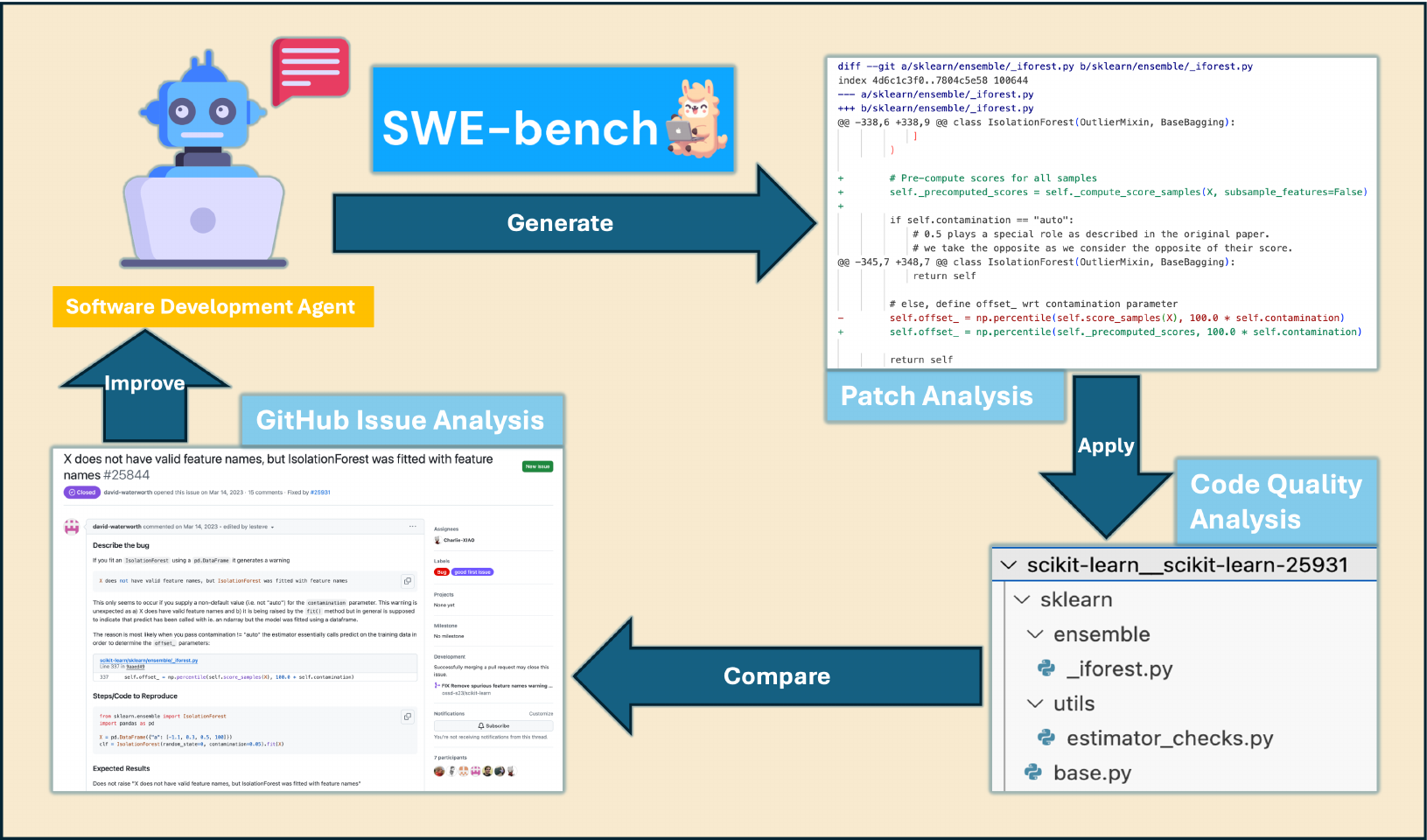}
    \caption{Overview: we evaluate agent-generated patches on SWE-Bench tasks, analyze their impact on the codebase, and compare resolved and unresolved GitHub issues to gain insights for improving agent performance.}
    \label{fig:overview}
\end{figure}
\vspace{-0.5em}

\textit{\textbf{Objectives:}} Figure \ref{fig:overview} presents an overview of our study, which aims to comprehensively evaluate software development agents' patch solutions for real-world GitHub issues from the \textit{SWE-Bench Verified} dataset~\cite{jimenez2024swebench}. First, we explore the patterns of agent-generated patches by comparing them to the \textit{gold patches} created by official repository developers. This comparison highlights the different approaches agents take to resolve issues, focusing on variations in file, function, and line-level modifications for the same problems. Next, we assess the broader impact of these patches on code quality, examining whether they introduce or resolve code smells, vulnerabilities, bugs, complexity, and duplication. Finally, we compare resolved and unresolved Github issues, identifying factors like problem statement complexity, codebase size, and solution effort that may affect agent performance. These insights offer practical recommendations to improve agent effectiveness in real-world settings.

\textit{\textbf{Main Contributions:}}
\begin{itemize}[noitemsep,nolistsep,leftmargin=1em]
    \item To the best of our knowledge, this is the first study to evaluate the quality of software development agents' generated patches for real-world GitHub issues.
    \item We analyze the reliability, security, and maintainability of agent-generated patches compared to human-written patches.
    \item We identify limitations in SWE-bench, as its unit tests do not fully cover all modified parts due to the diversity of agent-generated solutions.
    \item By comparing resolved and unresolved issues, we highlight their differences and offer suggestions for improving agent performance on more complex real-world tasks.
    \item To facilitate further research and enable reproducibility, we publicly share our datasets and scripts.\footnote{\url{https://osf.io/5urgc/?view_only=210932a785204432b86d857c089e25dd}}
\end{itemize}


\section{Study Design}

\subsection{\textbf{Choice of Benchmark}} 
We aim to evaluate the quality of agent-generated patches in real-world scenarios. For this, we use the SWE-Bench Verified dataset~\cite{chowdhury2024introducing}, which contains 500 Issue–Pull Request pairs from 12 open-source Python repositories. Validated by software engineers with support from OpenAI's Preparedness Team, it offers a reliable benchmark for assessing agents on real GitHub issues. Each issue is tied to a PR containing solution code and unit tests based on \textit{gold patches} from the repository developers. These tests include \texttt{FAIL\_TO\_PASS} tests, which fail before the PR and pass after, verifying that the issue is resolved, and \texttt{PASS\_TO\_PASS} tests, which pass both before and after, ensuring that unrelated functionality remains intact. Agents are given the issue text (problem statement) and access to the codebase but not the tests. An issue is considered \texttt{RESOLVED} if the agent's code passes both test types, ensuring the solution is correct and does not break existing functionality. This evaluation framework, coupled with a public leaderboard tracking agent performance, provides a robust benchmark for assessing agent-generated patches~\cite{jimenez2024swebench}.

\subsection{\textbf{Choice of Agents}}
We selected the top 10 agents from the SWE-Bench Verified public leaderboard~\footnote{\url{https://www.swebench.com/}} as of August 25, 2024, representing both industry and academia. These agents, recognized for their high performance in resolving GitHub issues, represent the state-of-the-arts in the latest advancements in AI-driven software development. By evaluating this diverse group, we provide a comprehensive assessment of the quality of their generated patches. The rankings and data reflect the most current results, ensuring the relevance and accuracy of our evaluation. Table \ref{tab:top_agents} presents the details and reported issue resolution rates for these agents.


\begin{table}[h]
    \centering
    \caption{Top 10 Agents from SWE-bench Verified Leaderboard}
    \label{tab:top_agents}
    \resizebox{\columnwidth}{!}{%
    \begin{tabular}{l l r r r}
        \toprule
        \textbf{Rank} & \textbf{Agents} & \textbf{Org Type} & \textbf{\% Resolved} & \textbf{Date} \\
        \midrule
        1 & Gru & Industry & 45.20\% & 24-08-24 \\
        2 & HoneyComb & Industry & 40.60\% & 24-08-20 \\
        3 & Amazon Q Developer Agent (v20240719) & Industry & 38.80\% & 24-07-21 \\
        4 & AutoCodeRover (v20240620) + GPT 4o & Academia & 38.40\% & 24-06-28 \\
        5 & Factory Code Droid & Industry & 37.00\% & 24-06-17 \\
        6 & SWE-agent + Claude 3.5 Sonnet & Academia & 33.60\% & 24-06-20 \\
        7 & AppMap Navie + GPT 4o & Industry & 26.20\% & 24-06-15 \\
        8 & Amazon Q Developer Agent (v20240430) & Industry & 25.60\% & 24-05-09 \\
        9 & EPAM AI/Run Developer Agent + GPT4o & Industry & 24.00\% & 24-08-20 \\
        10 & SWE-agent + GPT 4o & Academia & 23.20\% & 24-07-28 \\
        \bottomrule
    \end{tabular}
    }
\end{table}

\subsection{\textbf{Research Questions}}
\label{sec:RQs}

\textit{\textbf{RQ1:}What patch patterns do current Software Development Agents use when solving real-world GitHub issues?}


\textit{Motivation:} The SWE-Bench Verified dataset presents complex tasks where agents must analyze problem statements, identify relevant files in large codebases, and generate patches to resolve issues. While the current leaderboard only measures the percentage of issues resolved, it lacks deeper 
analysis 
into how agents generate these patches. Our goal is to go beyond this basic metric by comparing agent-generated patches to human-developed gold patches, exploring whether agents modify similar files and functions or 
make alternative modifications.
This will help uncover how closely agent solutions align with human solutions and reveal the nuances of their patch generation.

\textit{{\textbf{RQ2:} How do patches generated by Software Development Agents impact the reliability, security, and maintainability of the codebase?}}

\textit{Motivation:} The current SWE-Bench Verified benchmark focus on
passing the given test cases but overlook how agent-generated patches affect other aspects
like overall reliability, security, and maintainability. Solely evaluating issue resolution
on limited test cases
can miss broader implications~\cite{reis2021fixing}. Our goal is to assess whether these patches introduce or resolve code smells~\cite{tufano2017and}, vulnerabilities, bugs, increase code complexity~\cite{peitek2021program}, or duplication~\cite{mo2023comprehensive}. This deeper evaluation provides a more
comprehensive
understanding of their impact on overall software quality.

\textit{{\textbf{RQ3:} What differentiates resolved and unresolved GitHub issues, and how can these
differences
be used to improve the Issue Resolved Rate of Software Development Agents?}}

\textit{Motivation:} Despite progress, a significant number of GitHub issues in the SWE-Bench Verified dataset remain unresolved. To explore the differences between resolved and unresolved issues, we conduct an in-depth comparative analysis, focusing on factors like problem statement readability~\cite{xiao2022recommending}, codebase size, and solution effort. This analysis provides
insights into the challenges agents still face, offering practical recommendations to enhance their success in real-world 
settings.

\section{Data Collection and Construction}

Our evaluation data consists of mainly three portions: the official \textit{SWE-Bench Verified} dataset, \textit{agent-generated patch solutions}, and the \textit{code files associated with each patch}.
The following subsections introduce these datasets, which are the bases for our later analyses.

\subsection{\textbf{SWE-Bench Verified Dataset}}
We downloaded the \textit{SWE-Bench Verified} dataset from Hugging Face\footnote{\url{https://huggingface.co/datasets/princeton-nlp/SWE-bench_Verified}}. 
It includes 500 human-validated samples from the larger SWE-Bench dataset of 2,200 samples. 
Each sample in the dataset has been reviewed for quality by OpenAI's Preparedness Team~\cite{chowdhury2024introducing}. The key components in this dataset are: 1) \texttt{repo} - The repository owner/name identifier from GitHub; 
2) \texttt{base\_commit} - The commit
before the solution
patch
is applied; 
3) \texttt{problem\_statement} - The issue title and body describing the problem; 
4) \texttt{patch} - The gold patch created by repository developers to resolve the issue.


\subsection{\textbf{Agent-Generated Patches}}
For each agent, we collected the agent's patch solutions from the SWE-Bench Verified public leaderboard by extracting from its logs and prediction files (e.g., \texttt{all\_preds.jsonl}) for the 500 GitHub issues. The generated \texttt{patch.diff} files represent the agent's attempts to resolve these issues.



\begin{table}[h]
    \centering
    \caption{Summary of Collected Patches and Code Files}
    \label{tab:patches_codes_summary}
    \resizebox{\columnwidth}{!}{%
    \begin{tabular}{l r r r r r}
        \toprule
        \textbf{Source} & \multicolumn{2}{c}{\textbf{Patches}} & \multicolumn{3}{c}{\textbf{Code Files}} \\
        \cmidrule(lr){2-3} \cmidrule(lr){4-6}
        & \textbf{Generated} & \textbf{Applied} & \textbf{Pre-patch} & \textbf{Post-patch} & \textbf{Difference} \\
        \midrule
        \rowcolor{gold} Gold (Repo Developers) & 500 & 500 & 621 & 622 & {\color{darkgreen}+1} \\
        \midrule
        Gru & 500 & 499 & 622 & 622 & 0 \\
        HoneyComb & 486 & 469 & 723 & 723 & 0 \\
        Amazon-Q-Dev\_v240719 & 499 & 499 & 563 & 563 & 0 \\
        AutoCodeRover\_GPT4o & 492 & 486 & 542 & 542 & 0 \\
        FactoryCodeDroid & 500 & 500 & 512 & 512 & 0 \\
        SWE-Agent\_Claude3.5 & 489 & 459 & 574 & 1214 & {\color{darkgreen}+640} \\
        AppMap-Navie\_GPT4o & 494 & 494 & 680 & 680 & 0 \\
        Amazon-Q-Dev\_v240430 & 500 & 498 & 548 & 549 & {\color{darkgreen}+1} \\
        EPAM-Dev\_GPT4o & 482 & 482 & 684 & 690 & {\color{darkgreen}+6} \\
        SWE-Agent\_GPT4o & 450 & 434 & 484 & 1062 & {\color{darkgreen}+578} \\
        \bottomrule
    \end{tabular}
    }
\end{table}

\subsection{\textbf{Patch-Associated Code Files}}

To assess the impact of the patches, we retrieved the pre-patch relevant files from the base commit of each repository and applied the corresponding \texttt{patch.diff} files to generate the post-patch files. This
processing
enabled us to track changes in code quality, the number of files, and modifications between the pre-patch and post-patch states. Table \ref{tab:patches_codes_summary} summarizes the \textit{number of patches generated} by each agent, the \textit{number of patches that were applied} without errors (regardless of whether they resolved the issue), and the \textit{total number of files before and after applying the patches}.

\label{sec:patch_analysis}
\section{Patch Analysis} 
\subsection{\textbf{Experimental Setup}}
To answer \textit{\textit{\textbf{RQ1:} What patch patterns do current Software Development Agents use when solving real-world GitHub issues?}} we designed a multi-level analysis to evaluate agent-generated patches. This approach begins at the issue level, where we assess overall problem-solving success, and drills down to file, function, and line-level modifications. The purpose of this structure is to progressively reveal how agents handle increasingly granular aspects of software development, allowing us to understand both high-level patching performance and detailed patch
patterns.

\textit{\textbf{Issue Level:}} At the issue level, our goal is to understand how effectively agents are resolving real-world GitHub issues. We categorize issues based on how many agents successfully resolved them—starting from those solved by all 10 agents to those solved by only one or none. This allows us to identify common challenges that agents consistently resolve and issues that remain unsolved by all agents. Additionally, we perform an overlap analysis to explore whether certain agents dominate specific issues or if there is complementary performance between agents. This helps reveal patterns of strength and weakness among the top-performing agents, shedding light on whether individual agents specialize in certain types of issues or if there is significant overlap in their success.



\textit{\textbf{File and Function Level:}} We analyze which files and functions each agent modifies compared to the repository developers. Although a solution may pass all test cases, these are based on gold patches, and if agents modify different parts of the code, the tests may not fully cover their changes. We assess whether agents target the same files and functions as developers, using \textit{precision}, \textit{recall}, and \textit{F1-score} to quantify alignment with gold patches~\cite{johansson2024mapping}. This helps determine how effectively agents identify relevant code areas and make targeted modifications.

\textit{\textbf{Line Level:}} At the line level, we assess the patterns of how agents modify code through metrics such as \textit{added lines}, \textit{deleted lines}, \textit{total edits} (sum of additions and deletions), and \textit{net code size change} (difference between added and deleted lines). These patterns are key for understanding how agents manage code complexity and maintainability. Excessive changes can introduce complexity, while minimal edits may leave issues unresolved~\cite{codecomplexity2017antinyan}. To quantify differences between agent and gold patch modifications, we apply the \textit{Wilcoxon signed-rank test}~\cite{karakativc2022software} to identify statistically significant variations in these line-level patterns.

\subsection{\textbf{Experimental Result}}

\textit{\textbf{Issue-Level Analysis:}} In this section, we explore the performance of agents on the issue level, focusing on how many issues were resolved and the overlaps between agents.

\begin{figure}[h]
    \centering
    \includegraphics[width=\columnwidth]
    {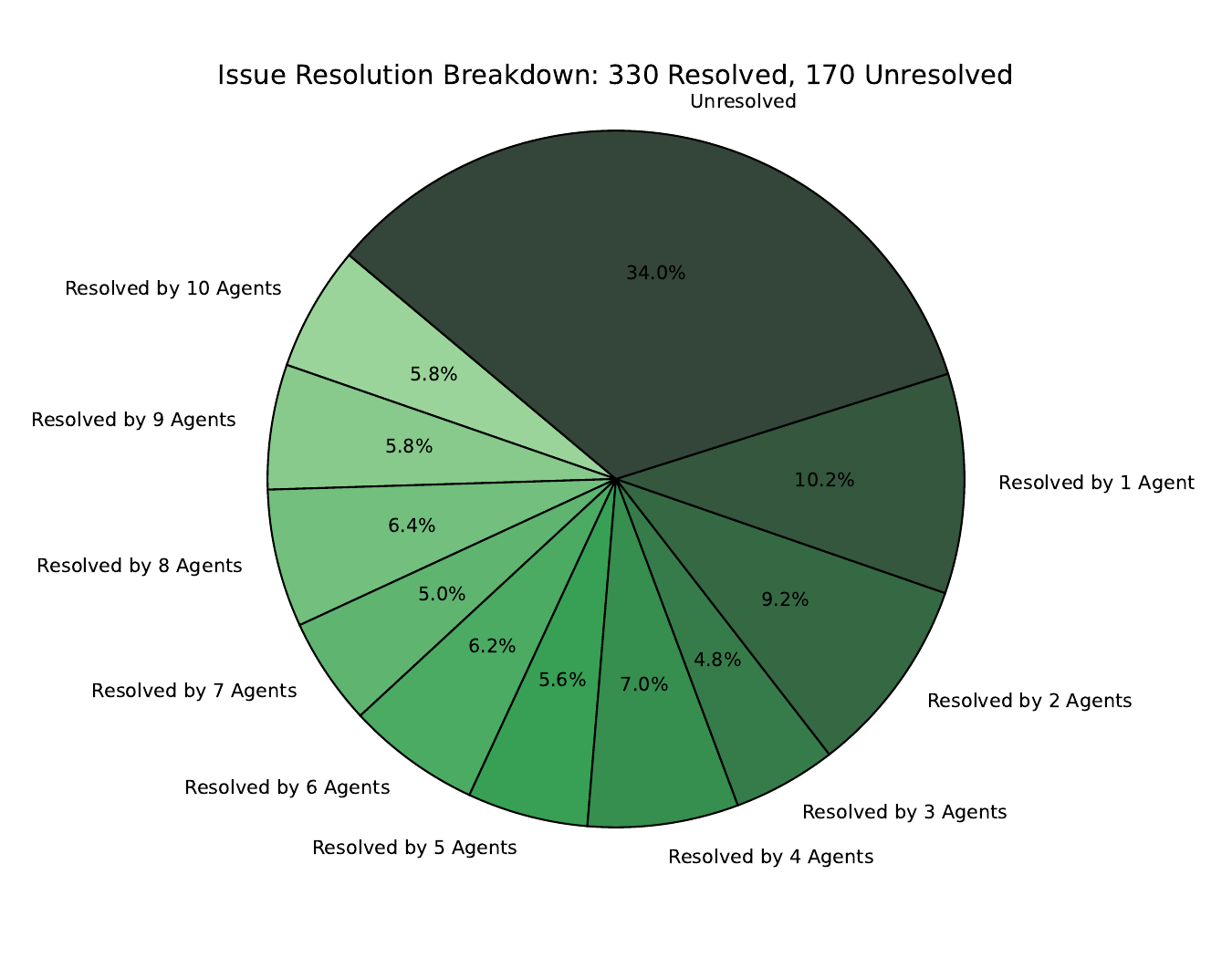}
    \caption{Issue Resolution Breakdown: 330 Resolved, 170 Unresolved}
    \label{fig:issue_pie}
\end{figure}

\textit{Issue Resolution Breakdown:} Figure~\ref{fig:issue_pie} shows the breakdown of resolved and unresolved issues across the top 10 agents. A total of 500 issues were analyzed, with 330 issues resolved and 170 remaining unresolved. Notably, 34\% of issues remain unresolved, indicating that a substantial portion of the issues could not be addressed by any one of the top 10 agents.  On the other hand, only 10.2\% of the issues were resolved by a single agent, and a relatively small proportion (5.8\%) were solved by all 10 agents. This suggests that while some issues are universally solvable, many are more specialized, requiring unique capabilities from different agents. The diversity of agent strengths highlights the complementary roles these agents play in resolving GitHub issues.

\begin{figure}[h]
    \centering
    \includegraphics[width=\columnwidth]{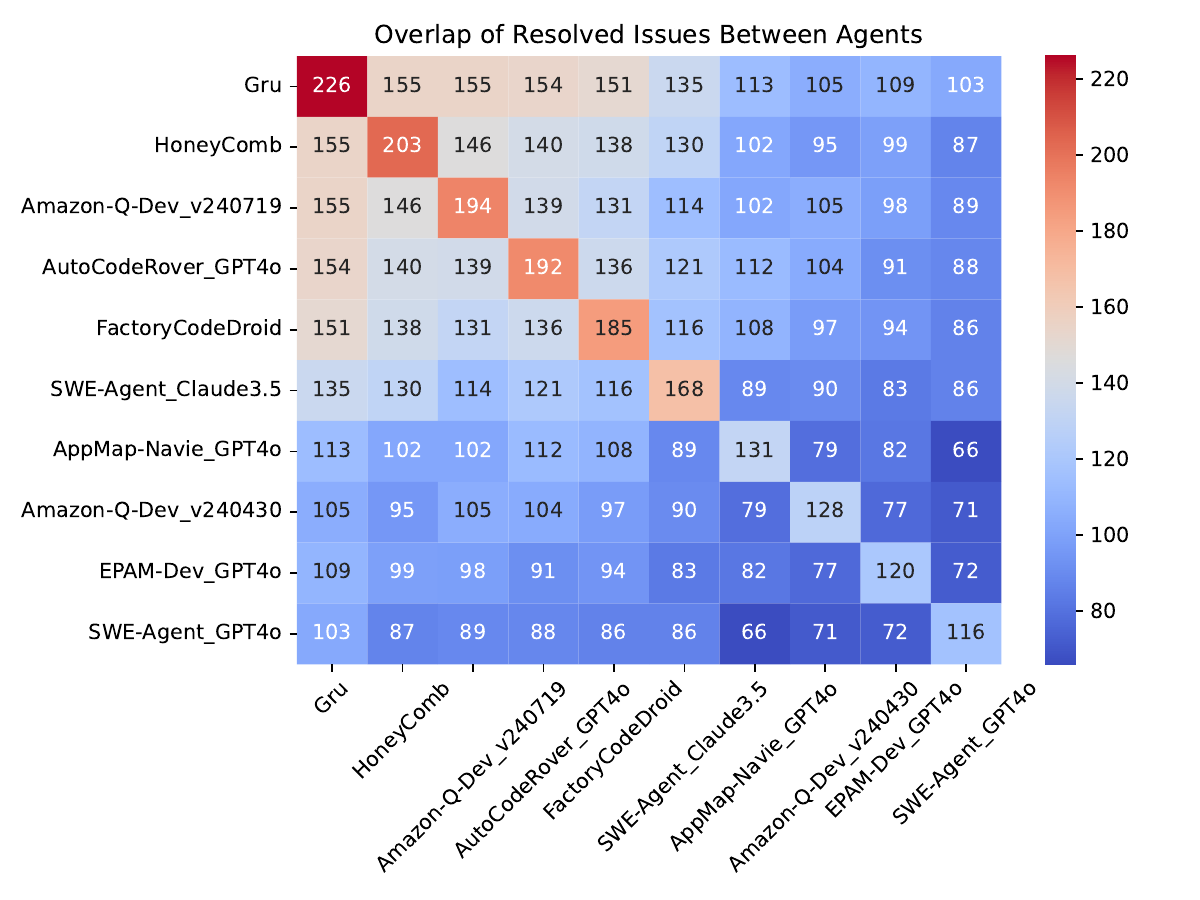}
    \caption{Overlap of Resolved Issues Between Agents}
    \label{fig:overlap_heatmap}
\end{figure}

\textit{Overlap of Resolved Issues Between Agents:} Figure~\ref{fig:overlap_heatmap} illustrates the overlap of resolved issues between agents. Each cell shows the number of issues resolved by both agents, with the diagonal representing the total issues each agent resolved. While \textit{Gru}, which resolved 226 issues, shares some overlap with other high-ranking agents like \textit{Honeycomb} (203 issues with 155 overlapping), even though \textit{Honeycomb} has a significant overlap with \textit{Gru}, a substantial number of issues (48 out of 203) remain unique. Similarly, relatively lower-ranking agents such as \textit{SWE-Agent\_Claude3.5}, which resolved 168 issues, had 33 that were unique compared to \textit{Gru}. Another example is \textit{Appmap-Navie\_GPT4o}, which resolved 131 issues, with 113 overlapping and 18 being unique compared to \textit{Gru}. The overall results demonstrate that no single agent covers all the issues resolved by others, indicating that all agents can learn from each other. While top agents like \textit{Gru} perform well overall, they can still benefit from the unique solutions offered by other agents, as many resolved issues are not shared.

\begin{tcolorbox}[colback=yellow!10!white, colframe=yellow!40!black, title=Finding 1]
No single agent dominates, as each can potentially learn from others to cover cases they currently miss. 170 issues remain unresolved, emphasizing the need for further improvements in agent capabilities.
\end{tcolorbox}

\begin{table*}[htbp]
    \centering
    \caption{Resolved and Unresolved GitHub Issues: F1 Scores}
    \resizebox{\textwidth}{!}{%
    \begin{tabular}{lcrrrrcrrrr}
        \toprule
        \multirow{3}{*}{\textbf{Agent}} & \multicolumn{5}{c}{\textbf{Resolved Issues}} & \multicolumn{5}{c}{\textbf{Unresolved Issues}} \\
        \cmidrule(lr){2-6} \cmidrule(lr){7-11}
         &  & \multicolumn{2}{c}{\textbf{Files}} & \multicolumn{2}{c}{\textbf{Functions}} &  & \multicolumn{2}{c}{\textbf{Files}} & \multicolumn{2}{c}{\textbf{Functions}} \\
         
        \cmidrule(lr){3-4} \cmidrule(lr){5-6}  \cmidrule(lr){8-9} \cmidrule(lr){10-11}
        
         & \textbf{Total} & \textbf{Total F1=1} & \textbf{Ratio F1=1 (\%)} & \textbf{Total F1=1} & \textbf{Ratio F1=1 (\%)} & \textbf{Total} & \textbf{Total F1=1} & \textbf{Ratio F1=1 (\%)} & \textbf{Total F1=1} & \textbf{Ratio F1=1 (\%)} \\
        \midrule
        Gru & 226 & 197 & 87.17\% & 55 & 24.34\% & 274 & 124 & 45.26\% & 18 & 6.57\% \\
        HoneyComb & 203 & 111 & 54.68\% & 30 & 14.78\% & 283 & 79 & 27.92\% & 10 & 3.53\% \\
        Amazon-Q-Dev\_v240719 & 194 & 163 & 84.02\% & 12 & 6.19\% & 305 & 146 & 47.87\% & 3 & 0.98\% \\
        AutoCodeRover\_GPT4o & 192 & 171 & 89.06\% & 50 & 26.04\% & 300 & 149 & 49.67\% & 15 & 5.00\% \\
        FactoryCodeDroid & 185 & 175 & 94.59\% & 45 & 24.32\% & 315 & 173 & 54.92\% & 32 & 10.16\% \\
        SWE-Agent\_Claude3.5 & 168 & 26 & 15.48\% & 22 & 13.10\% & 321 & 20 & 6.23\% & 10 & 3.12\% \\
        AppMap-Navie\_GPT4o & 131 & 108 & 82.44\% & 29 & 22.14\% & 363 & 130 & 35.81\% & 26 & 7.16\% \\
        Amazon-Q-Dev\_v240430 & 128 & 115 & 89.84\% & 18 & 14.06\% & 372 & 192 & 51.61\% & 9 & 2.42\% \\
        EPAM-Dev\_GPT4o & 120 & 78 & 65.00\% & 26 & 21.67\% & 362 & 122 & 33.70\% & 20 & 5.52\% \\
        SWE-Agent\_GPT4o & 116 & 70 & 60.34\% & 29 & 25.00\% & 334 & 71 & 21.26\% & 31 & 9.28\% \\
        \bottomrule
    \end{tabular}%
    }
    \label{tab:f1_scores}
\end{table*}


\textit{\textbf{Files and Functions Level:}} The results in Table~\ref{tab:f1_scores} show differences between agent-generated patches and the gold patches in SWE-bench. While multiple valid solutions exist, the table highlights instances where \textit{F1=1}. Since the F1-score is the harmonic mean of precision and recall, an F1=1 means both precision and recall are also 1, indicating that the agent made the exact same file or function changes as the gold patch. In such cases, the agent's modifications align perfectly with the gold patch, meaning these cases are well-covered and evaluated by the existing test cases.

For resolved issues, agents such as \textit{Gru} and \textit{FactoryCodeDroid} demonstrate high precision in modifying the same files as the gold patches, with ratios of \textit{87.17\%} and \textit{94.59\%}, respectively. However, accuracy at the function level is notably lower, with ratios of \textit{24.34\%} and \textit{24.32\%}, indicating that while agents are often identifying the correct files, they may be modifying different functions compared to the gold patch solutions. For unresolved issues, agents like \textit{AppMap-Navie\_GPT4o} and \textit{SWE-Agent\_Claude3.5} show even lower alignment with the gold patches at both the file and function levels, potentially contributing to their lower success rates in resolving issues.

A core concern is that SWE-bench uses \textit{unit tests} to verify patches, including \texttt{FAIL\_TO\_PASS} tests to ensure the issue is resolved, and \texttt{PASS\_TO\_PASS} tests to confirm unrelated functionality remains intact. However, since these tests are based on the gold patches' modifications, agent-generated patches that alter different files or functions may not be fully covered, risking other parts of the codebase being broken despite passing all tests.

\begin{tcolorbox}[colback=yellow!10!white, colframe=yellow!40!black, title=Finding 2] 
The experiment reveals a limitation in SWE-bench's evaluation. Agent-generated patches, though passing unit tests, may break other functionalities by modifying different files and functions than the gold patches, a risk not fully captured by current test coverage. \end{tcolorbox}

\textit{\textbf{Line Level Analysis:}} Table~\ref{tab:resolved_lines_changes_comparison} highlights differences between agent-generated patches and gold patches for resolved issues, focusing on total code changes and net code size changes.

\begin{table}[htbp]
    \centering
    \caption{Comparison of Total Changes and Net Code Size Changes for Resolved Instances}
    \resizebox{\columnwidth}{!}{%
    \begin{tabular}{lrrrr|rrrrr}
        \toprule
        \multirow{2}{*}{\textbf{Agent}} & \multicolumn{3}{c}{\textbf{Total Code Changes}} & & \multicolumn{3}{c}{\textbf{Net Code Size Changes}} \\
        \cmidrule(lr){2-4} \cmidrule(lr){6-8}
         & \textbf{Agent Mean} & \textbf{Gold Mean} & \textbf{Significant} & & \textbf{Agent Mean} & \textbf{Gold Mean} & \textbf{Significant} \\
        \midrule
        Gru                 & 5.67  & 7.48   & \textcolor{darkgreen}{\checkmark}  & & 2.32  & 2.16  & \textcolor{red}{\texttimes} \\
        HoneyComb           & 47.54 & 7.71   & \textcolor{darkgreen}{\checkmark}  & & 34.79 & 2.36  & \textcolor{darkgreen}{\checkmark}  \\
        Amazon-Q-Dev\_v240719   & 8.42  & 7.42   & \textcolor{red}{\texttimes}  & & 3.23  & 2.34  & \textcolor{darkgreen}{\checkmark}  \\
        AutoCodeRover\_GPT4o    & 6.54  & 7.92   & \textcolor{darkgreen}{\checkmark}  & & 2.78  & 2.19  & \textcolor{darkgreen}{\checkmark}  \\
        FactoryCodeDroid & 6.98  & 7.55   & \textcolor{red}{\texttimes}  & & 3.63  & 2.25  & \textcolor{darkgreen}{\checkmark}  \\
        SWE-Agent\_Claude3.5  & 43.35 & 7.85   & \textcolor{darkgreen}{\checkmark}  & & 27.83 & 1.57  & \textcolor{darkgreen}{\checkmark}  \\
        AppMap-Navie\_GPT4o  & 8.12  & 6.89   & \textcolor{darkgreen}{\checkmark}  & & 3.01  & 2.36  & \textcolor{darkgreen}{\checkmark}  \\
        Amazon-Q-Dev\_v240430   & 6.08  & 7.91   & \textcolor{darkgreen}{\checkmark}  & & 1.56  & 1.74  & \textcolor{red}{\texttimes} \\
        EPAM-Dev\_GPT4o          & 10.87 & 6.53   & \textcolor{darkgreen}{\checkmark}  & & 7.57  & 1.63  & \textcolor{darkgreen}{\checkmark}  \\
        SWE-Agent\_GPT4o      & 21.67 & 7.43   & \textcolor{darkgreen}{\checkmark}  & & 17.26 & 0.93  & \textcolor{darkgreen}{\checkmark}  \\
        \bottomrule
    \end{tabular}%
    }
    \label{tab:resolved_lines_changes_comparison}
\end{table}

Several agents, such as \textit{Gru} and \textit{Amazon-Q-Dev\_v240719}, align closely with the gold patches in terms of net code size changes, suggesting they modify the code similarly to the gold patches in terms of overall impact. However, agents like \textit{HoneyComb} and \textit{SWE-Agent\_Claude3.5} show significant deviations, indicating a tendency to either over-modify or under-modify the code, which can affect maintainability by introducing complexity or leaving issues unresolved. Agents like \textit{AppMap-Navie\_GPT4o} and \textit{FactoryCodeDroid} differ in total changes but align closely in net code size, suggesting alternative approaches that achieve similar overall impact. The results also show that some agents, such as \textit{SWE-Agent\_GPT4o}, significantly increase code size, potentially posing challenges for long-term maintainability, an issue further analyzed in Section~\ref{sec:code_analysis}.

\begin{tcolorbox}[colback=yellow!10!white, colframe=yellow!40!black, title=Finding 3]
The line-level analysis reveals that agents like \textit{HoneyComb} and \textit{SWE-Agent\_Claude3.5} tend to over-modify the code, leading to significant increases in net code size.
In contrast, agents like \textit{Gru} and \textit{Amazon-Q-Dev\_v240719} demonstrate closer alignment with the gold patches, showing more balanced modifications.
\end{tcolorbox}

\section{Code Quality Analysis}
\label{sec:code_analysis}

\subsection{\textbf{Experimental Setup}}

To answer \textit{\textbf{RQ2:} How do patches generated by Software Development Agents impact the reliability, security, and maintainability of the codebase?} we focus on each agent's patches that have successfully \textbf{\textsc{resolved}} issues—those most likely to be accepted and merged into the codebase. Given their potential to impact the codebase long-term, it is crucial to evaluate these patches beyond functional correctness. For each agent’s group of resolved patches, we assess their impact across three key non-functional aspects: \textit{reliability}, \textit{security}, and \textit{maintainability}. These aspects are essential for understanding the broader effects on code quality and sustainability.
We use \textit{SonarQube} to perform static analysis for these metrics, as it is widely recognized in both academic research and industry~\cite{nguyen2022empirical,liu2024no,peitek2021program}. Its free community version ensures the reproducibility of our experiments, while sharing the same set of detection rules as the commercial version used in industry. This alignment guarantees that our findings are both accessible and relevant to real-world software quality concerns.

\paragraph{\textbf{Reliability}} 
The primary goal of assessing reliability is to determine whether the agent-generated patches introduce new bugs or fix existing ones without breaking other parts of the codebase~\cite{falcao2020relating}. This is crucial because a patch, while solving one problem, could inadvertently destabilize other parts of the system. To evaluate this, we measure the number of bugs in the pre-patch files and compare them to the post-patch files using SonarQube’s bug detection capabilities. This approach provides insight into whether agents maintain or degrade the overall stability of the code.

\paragraph{\textbf{Security}}
Security is crucial in software development, as introducing vulnerabilities can lead to severe consequences~\cite{zhou2021finding}. To assess whether agent-generated patches improve or weaken the security of the codebase, we use SonarQube to calculate the number of vulnerabilities in the pre-patch files and then re-evaluate them after the patch is applied. This analysis helps determine whether the agent patches not only address the issue but also avoid creating new security risks. Given the increasing importance of secure software, this step ensures that agent-generated patches contribute positively to the overall security posture of the software.

\paragraph{\textbf{Maintainability}}
Maintainability measures how easily the code can be understood, modified, and extended in the future, which is essential for the long-term sustainability of software~\cite{sharma2021qscored}. We use SonarQube to evaluate three key metrics: code smells, code complexity, and code duplication.

\begin{itemize}[leftmargin=1em]
    \item \textbf{Code complexity}: We measure code cyclomatic complexity, which quantifies the number of independent paths through a function’s control flow~\cite{yan2023closer}. Since highly complex code is harder to modify and maintain, we normalize this value by dividing the cyclomatic complexity by the number of lines of code to give a balanced view of code complexity relative to its size.
    \item \textbf{Code duplication}: Duplicated code increases maintenance costs, as changes must be applied in multiple places. We assess the percentage of duplication by calculating the ratio of duplicate lines to the total lines of code, providing insight into the risk of code bloat and unnecessary repetition.
    
    \item \textbf{Code smells}: Indicators of potential design flaws that can make the code harder to maintain. We calculate the total number of code smells in the pre-patch and post-patch files and normalize them by lines of code, giving a clearer picture of their prevalence relative to the size of the codebase.
\end{itemize}

By comparing these metrics before and after the patch is applied, we can determine whether the agent patches improve or worsen the maintainability of the code. Since the data for these metrics does not follow a normal distribution, as verified by the \textit{Shapiro-Wilk test}~\cite{yap2011comparisons}, we use the non-parametric \textit{Wilcoxon signed-rank test}~\cite{karakativc2022software} to determine whether there are statistically significant improvements or declines in code quality after the patches are applied. Additionally, we compute the \textit{Rank-Biserial Correlation}~\cite{cureton1956rank} to quantify the magnitude of changes, interpreting effect sizes using Cohen's guidelines~\cite{zhang2024assessing}.


\subsection{\textbf{Code Reliability Results}}
Table \ref{tab:bugs_comparison} compares the pre- and post-patch bug counts for each agent and the gold patches in resolved issues. This analysis focuses on whether agents were able to resolve issues without introducing new bugs.


\begin{table}[htbp]
    \centering
    \caption{Bug Count Analysis (Resolved Patches)}
    \resizebox{\columnwidth}{!}{%
    \begin{tabular}{lrrrrr}
        \toprule
        \textbf{Patch Source} & \textbf{Pre-patch Mean} & \textbf{Post-patch Mean} & \textbf{Significance} & \textbf{Effect size} & \textbf{Effect size interpretation} \\
        \midrule
         \rowcolor{gold}Gold & (97/500) 0.1940 & (95/500) 0.1900 & \textcolor{red}{\xmark} & -1.0000 & \cellcolor{red!30}Large \\
         \midrule
        Gru & (28/226) 0.1239 & (35/226) 0.1549 & \textcolor{red}{\xmark} & 0.7500 & \cellcolor{red!30}Large \\
        HoneyComb & (62/188) 0.3298 & (61/188) 0.3245 & \textcolor{red}{\xmark} & 0.0000 & Negligible \\
        Amazon-Q-Dev\_v240719 & (28/194) 0.1443 & (26/194) 0.1340 & \textcolor{red}{\xmark} & -1.0000 & \cellcolor{red!30}Large \\
        AutoCodeRover\_GPT4o & (25/191) 0.1309 & (24/191) 0.1257 & \textcolor{red}{\xmark} & 0.0000 & Negligible \\
        FactoryCodeDroid & (32/185) 0.1730 & (33/185) 0.1784 & \textcolor{red}{\xmark} & 0.5000 & \cellcolor{red!30}Large \\
        SWE-Agent\_Claude3.5 & (37/165) 0.2242 & (37/165) 0.2242 & \textcolor{red}{\xmark} & 0.3333 & Medium \\
        AppMap-Navie\_GPT4o & (18/131) 0.1374 & (18/131) 0.1374 & \textcolor{red}{\xmark} & 0.3333 & Medium \\
        Amazon-Q-Dev\_v240430 & (18/128) 0.1406 & (19/128) 0.1484 & \textcolor{red}{\xmark} & 0.5000 & \cellcolor{red!30}Large \\
        EPAM-Dev\_GPT4o & (45/120) 0.3750 & (43/120) 0.3583 & \textcolor{red}{\xmark} & 0.0000 & Negligible \\
        SWE-Agent\_GPT4o & (12/114) 0.1053 & (13/114) 0.1140 & \textcolor{red}{\xmark} & 0.3333 & Medium \\
        \bottomrule
    \end{tabular}%
    }
    \begin{flushleft}
    \textbf{Note:} values in parentheses show the \textit{total bugs identified} by SonarQube and \textit{total issues resolved} by each agent. 
    



    \end{flushleft}
    \label{tab:bugs_comparison}
\end{table}

\textit{Interpretation:} Overall, most agents maintained reliability by resolving issues without introducing significant new bugs. Some agents, like \textit{Gru} and \textit{FactoryCodeDroid}, show minor increases in bug count, though none of these changes are statistically significant. In contrast, agents such as \textit{Amazon-Q-Dev\_v240719} and \textit{EPAM-Dev\_GPT4o} demonstrate decreases in bug count, suggesting potential improved reliability. Other agents, like \textit{HoneyComb} and \textit{AutoCodeRover\_GPT4o}, exhibit negligible changes, indicating no new bugs were introduced. While agents like \textit{SWE-Agent\_Claude3.5} and \textit{AppMap-Navie\_GPT4o} show medium effect sizes, the changes remain insignificant. Overall, most agents resolved issues without introducing significant new bugs, thereby maintaining code reliability.

\begin{tcolorbox}[colback=yellow!10!white, colframe=yellow!40!black, title=Finding 4]
Most agents resolved issues effectively without introducing significant new bugs, maintaining code reliability. However, \textit{Gru} and \textit{FactoryCodeDroid} showed slight increases in bug counts, though these changes were not statistically significant.
\end{tcolorbox}

\subsection{\textbf{Code Security Results}}
In this experiment, we assessed whether agent-generated patches introduced new vulnerabilities into the codebase when resolving GitHub issues. We evaluated both pre-patch and post-patch code files for vulnerabilities across all agents, as well as the gold patches developed by repository maintainers.

The experiment results indicate that \textit{0 vulnerabilities} were found in either the pre-patch or post-patch files for any patches, regardless of the source. This applies to both the gold patches and the agent-generated patches. While this demonstrates that no vulnerabilities were present or introduced in these specific GitHub scenarios, prior studies~\cite{pearce2022asleep,asare2024user,asare2023github,majdinasab2024assessing,hamer2024just} have shown that base Large Language Models (LLMs) can generate vulnerable code. However, our results suggest that in these cases, agent-generated patches did not introduce new vulnerabilities, indicating good performance in terms of security. Future research should explore agents' performance in vulnerability-prone scenarios to better assess their security impact~\cite{tony2023llmseceval,siddiq2022securityeval}.

\begin{tcolorbox}[colback=yellow!10!white, colframe=yellow!40!black, title=Finding 5]
Our experiment results indicate that no vulnerabilities were introduced by either agent-generated or gold patches in these GitHub issues.
\end{tcolorbox}

\subsection{\textbf{Code Maintainability Results}}
\subsubsection{\textbf{Code Complexity}}
The results in Table~\ref{tab:complexity_comparison} summarize the changes in code complexity after patches were applied, represented as the ratio of cyclomatic complexity to total lines of code.

\begin{table}[htbp]
    \centering
    \caption{Code Complexity Analysis - Ratio of Cyclomatic Complexity to Total Lines of Code}
    \resizebox{\columnwidth}{!}{%
    \begin{tabular}{lrrrrr}
        \toprule
        \textbf{Patch Source} & \textbf{Pre-patch Mean} & \textbf{Post-patch Mean} & \textbf{Significance} & \textbf{Effect Size} & \textbf{Effect Size Interpretation} \\
        \midrule
         \rowcolor{gold}Gold & 0.2691 & 0.2698 & \textcolor{red}{\xmark} & -0.0177 & Negligible \\
           \midrule
        Gru & 0.2632 & 0.2637 & \textcolor{darkgreen}{\checkmark} & 0.2625 & Small \\
        HoneyComb & 0.2413 & 0.2417 & \textcolor{red}{\xmark} & 0.0494 & Negligible \\
        Amazon-Q-Dev\_v240719 & 0.2590 & 0.2599 & \textcolor{darkgreen}{\checkmark} & 0.1565 & Small \\
        AutoCodeRover\_GPT4o & 0.2656 & 0.2661 & \textcolor{darkgreen}{\checkmark} & 0.3014 & Small \\
        FactoryCodeDroid & 0.2612 & 0.2615 & \textcolor{darkgreen}{\checkmark} & 0.2555 & Small \\
        SWE-Agent\_Claude3.5 & 0.2426 & 0.2357 & \textcolor{darkgreen}{\checkmark} & -0.5302 & \cellcolor{red!30}Large \\
        AppMap-Navie\_GPT4o & 0.2606 & 0.2611 & \textcolor{darkgreen}{\checkmark} & 0.1875 & Small \\
        Amazon-Q-Dev\_v240430 & 0.2611 & 0.2611 & \textcolor{red}{\xmark} & 0.1494 & Small \\
        EPAM\_GPT4o & 0.2314 & 0.2316 & \textcolor{red}{\xmark} & 0.0435 & Negligible \\
        SWE-Agent\_GPT4o & 0.2517 & 0.2464 & \textcolor{darkgreen}{\checkmark} & -0.0930 & Negligible \\
        \bottomrule
    \end{tabular}%
    }
    \label{tab:complexity_comparison}
\end{table}

\textit{Interpretation:} The \textit{gold patches} show negligible changes in complexity, with no statistically significant difference between the pre-patch and post-patch scores. This suggests that human-created patches maintained the original structure of the code without significantly altering its complexity.

In contrast, agents like \textit{Gru}, \textit{Amazon-Q-Dev\_v240719}, and \textit{AutoCodeRover\_GPT4o} show small but statistically significant increases in complexity after their patches. Although these increases are small, they imply that agent-generated patches may slightly increase the complexity of the code, potentially affecting its long-term maintainability. On the other hand, \textit{SWE-Agent\_Claude3.5} showed a notable reduction in complexity, indicating that its patches may have simplified the code.

Overall, most agents introduced only minor changes in complexity, and these small effect sizes suggest the patches did not drastically impact code complexity.

\begin{tcolorbox}[colback=yellow!10!white, colframe=yellow!40!black, title=Finding 6]
Most agents, such as \textit{Gru}, slightly increased the complexity of the code post-patch, though these changes were small. Notably, \textit{SWE-Agent\_Claude3.5} reduced code complexity, potentially improving code simplicity. However, overall changes in complexity were minimal across all agents.
\end{tcolorbox}

\subsubsection{\textbf{Code Duplication}}
Table~\ref{tab:duplication_ratio_comparison} presents the changes in code duplication, measured as the ratio of duplicated lines to total lines of code, for pre-patch and post-patch code files.

\begin{table}[htbp]
    \centering
    \caption{Code Duplication Analysis (Resolved Patches) - Ratio of duplicated lines to Total lines of code}
    \resizebox{\columnwidth}{!}{%
    \begin{tabular}{lrrrrr}
        \toprule
        \textbf{Patch Source} & \textbf{Pre-patch Mean} & \textbf{Post-patch Mean} & \textbf{Significance} & \textbf{Effect Size} & \textbf{Effect Size Interpretation} \\
        \midrule
         \rowcolor{gold}Gold & 0.0036 & 0.0036 & \textcolor{darkgreen}{\checkmark} & -0.7647 & Large \\
        \midrule
        Gru & 0.0032 & 0.0057 & \textcolor{red}{\xmark} & -0.4444 & Medium \\
        HoneyComb & 0.0055 & 0.0055 & \textcolor{darkgreen}{\checkmark} & -0.8947 & Large \\
        Amazon-Q-Dev\_v240719 & 0.0024 & 0.0024 & \textcolor{darkgreen}{\checkmark} & -0.7500 & Large \\
        AutoCodeRover\_GPT4o & 0.0035 & 0.0035 & \textcolor{darkgreen}{\checkmark} & -0.8462 & Large \\
        FactoryCodeDroid & 0.0024 & 0.0024 & \textcolor{darkgreen}{\checkmark} & -0.7500 & Large \\
        SWE-Agent\_Claude3.5 & 0.0034 & 0.0036 & \textcolor{darkgreen}{\checkmark} & -0.7143 & Large \\
        AppMap-Navie\_GPT4o & 0.0016 & 0.0032 & \textcolor{red}{\xmark} & -0.7143 & Large \\
        Amazon-Q-Dev\_v240430 & 0.0026 & 0.0026 & \textcolor{red}{\xmark} & -0.6000 & Large \\
        EPAM-Dev\_GPT4o & 0.0038 & 0.0040 & \textcolor{red}{\xmark} & -0.7778 & Large \\
        SWE-Agent\_GPT4o & 0.0031 & 0.0049 & \textcolor{red}{\xmark} & -0.1111 & Negligible \\
        \bottomrule
    \end{tabular}%
    }
    \label{tab:duplication_ratio_comparison}
\end{table}

\textit{Interpretation:} Overall, the effect sizes indicate that most agents either maintained or decreased their code duplication ratios, preserving maintainability. The \textit{Gold} patches, along with agents such as \textit{HoneyComb}, \textit{Amazon-Q-Dev\_v240719}, \textit{AutoCodeRover\_GPT4o}, and \textit{FactoryCodeDroid}, all showed large negative effect sizes, suggesting that they either prevented increases in duplicated code or reduced it, contributing to improved code structure. Although agents like \textit{Gru}, \textit{AppMap-Navie\_GPT4o}, and \textit{EPAM-Dev\_GPT4o} exhibited increases in duplication ratios, these changes were not statistically significant, indicating only slight risks of future maintenance challenges. Therefore, most agents demonstrate reliable performance in maintaining or improving code quality through reduced or stable code duplication.

\begin{tcolorbox}[colback=yellow!10!white, colframe=yellow!40!black, title=Finding 7]
The \textit{Gold} patches, along with most agents, either maintained or reduced code duplication levels, preserving maintainability. Agents such as \textit{Gru} and \textit{AppMap-Navie\_GPT4o} showed increases in duplication, but these were not statistically significant.

\end{tcolorbox}

\subsubsection{\textbf{Code Smells}}

Table~\ref{tab:code_smells_comparison} presents the changes in code smells, measured as the ratio of code smells to total lines of code, for pre-patch and post-patch code files.
\vspace{-2em}
\begin{table}[htbp]
    \centering
    \caption{Code Smells Analysis (Patches Resolved) - Ratio of Code Smells to Total Lines of Code}
    \resizebox{\columnwidth}{!}{%
    \begin{tabular}{lrrrrr}
        \toprule
        \textbf{Patch Source} & \textbf{Pre-patch Mean} & \textbf{Post-patch Mean} & \textbf{Significance} & \textbf{Effect Size} & \textbf{Effect Size Interpretation} \\
        \midrule
         \rowcolor{gold}Gold & 0.0188 & 0.0188 & \textcolor{darkgreen}{\checkmark} & -0.6077 & Large \\
           \midrule
        Gru & 0.0189 & 0.0188 & \textcolor{darkgreen}{\checkmark} & -0.7460 & Large \\
        HoneyComb & 0.0196 & 0.0194 & \textcolor{darkgreen}{\checkmark} & -0.6267 & Large \\
        Amazon-Q-Dev\_v240719 & 0.0189 & 0.0187 & \textcolor{darkgreen}{\checkmark} & -0.7377 & Large \\
        AutoCodeRover\_GPT4o & 0.0210 & 0.0208 & \textcolor{darkgreen}{\checkmark} & -0.8387 & Large \\
        FactoryCodeDroid & 0.0174 & 0.0177 & \textcolor{darkgreen}{\checkmark} & -0.6552 & Large \\
        SWE-Agent\_Claude3.5 & 0.0198 & 0.0194 & \textcolor{darkgreen}{\checkmark} & -0.6087 & Large \\
        AppMap-Navie\_GPT4o & 0.0193 & 0.0192 & \textcolor{darkgreen}{\checkmark} & -0.7561 & Large \\
        Amazon-Q-Dev\_v240430 & 0.0183 & 0.0179 & \textcolor{darkgreen}{\checkmark} & -0.7101 & Large \\
        EPAM\_Dev\_GPT4o & 0.0192 & 0.0189 & \textcolor{darkgreen}{\checkmark} & -0.7215 & Large \\
        SWE-Agent\_GPT4o & 0.0183 & 0.0182 & \textcolor{darkgreen}{\checkmark} & -0.4848 & Large \\
        \bottomrule
    \end{tabular}%
    }
    \label{tab:code_smells_comparison}
\end{table}
\vspace{-2em}


\textit{Interpretation:} Based on the effect sizes, all agents, including the \textit{gold patches}, demonstrate similar performance, with large negative effect sizes across the board. This suggests that both agent-generated and human-created patches effectively reduced the ratio of code smells, contributing to improved maintainability. The consistency in negative effect sizes across agents indicates that they are generally capable of matching human developers in reducing code smells. For \textit{FactoryCodeDroid}, while the mean ratio of code smells increased slightly, the large negative effect size suggests that this increase was driven by a few outlier patches. Despite these outliers, the overall trend shows a reduction in code smell ratio across most patches.

\begin{tcolorbox}[colback=yellow!10!white, colframe=yellow!40!black, title=Finding 8]
Most agents produced patches comparable to those of the repository developers, indicating their growing ability to match human-level code quality by avoiding the introduction of code smells.
\end{tcolorbox}

\label{sec:issue_analysis}
\section{GitHub Issue Analysis}
\subsection{\textbf{Experimental Setup}}
To address \textit{{\textbf{RQ3:} What differentiates resolved and unresolved GitHub issues, and how can these 
differences
be used to improve the Issue Resolved Rate of Software Development Agents?}} we systematically compare \textbf{\textsc{resolved}} GitHub issues (those successfully addressed by at least one agent) with \textbf{\textsc{unresolved}} issues.
This comparison is conducted through three key perspectives: (1) the complexity of the issue's \textit{problem statement}, (2) the \textit{source code files} associated with these issues (derived from gold patches), and (3) the \textit{gold patch solutions} provided by repository developers. These perspectives are chosen to comprehensively understand the multifaceted challenges that agents encounter when resolving issues.

\subsubsection{\textbf{Analysis of GitHub Problem Statements}}
Understanding the complexity of problem statements is crucial as it directly impacts an agent's ability to comprehend and address issues effectively. Therefore, we evaluate the problem statements using the following metrics:

\begin{itemize}[leftmargin=1em]
    \item \textit{Readability and Length:} We assess problem statement complexity using \textit{Flesch Reading Ease}~\cite{kincaid1975derivation}, where higher scores indicate easier text, and \textit{Flesch-Kincaid Grade Level}, which estimates the required education level. These metrics have been applied in previous work~\cite{xiao2022recommending,fan2018chaff,eleyan2020enhancing}. Additionally, we consider \textit{Sentence Count} and \textit{Word Count}, assuming that lower readability and longer content may hinder agents' understanding of the GitHub issue, making it more difficult to generate accurate solutions~\cite{li2022follow}.

    \item \textit{Code Relevance:} Code snippets in the problem statement can provide valuable context for agents, helping them locate and solve issues. However, they may also increase the difficulty for agents to process~\cite{zhang2024developers}. We measure metrics such as \textit{Contains Code snippets}, \textit{Number of Code Blocks}, \textit{Lines of Code}, and the \textit{Code-to-Text Ratio} to compare these factors between resolved and unresolved GitHub issue groups.
\end{itemize}

\subsubsection{\textbf{Analysis of Associated Source Code Files}}
Navigating, understanding, and modifying relevant source code files are crucial for resolving issues. We analyze the \textit{source code files} linked to each issue, using the \textit{gold patch}---the repository developers' solution---as the base for comparison:

\begin{itemize}[leftmargin=1em]
    \item \textit{Number of Modified Files:} A higher number of modified files typically indicates a more complex issue, requiring agents to handle changes across multiple parts of the codebase.
  
    \item \textit{Code Size and Cyclomatic Complexity:} We measure \textit{total lines of code} and \textit{cyclomatic complexity} to assess the difficulty agents face in understanding and generating effective patches. Larger codebases and higher complexity present greater challenges.
\end{itemize}

\subsubsection{\textbf{Analysis of Gold Patch Solutions}}
Gold patches provide indirect evidence of the scale and complexity of the changes needed to resolve issues. We assess these solutions using two key metrics:
\begin{itemize}[leftmargin=1em]
    \item \textit{Total Lines Change:} This measures the overall number of lines added and deleted, helping to determine the scale of modifications. Larger changes may indicate more complex restructuring.
  
    \item \textit{Net Code Size Change:} This metric evaluates whether the patch leads to an overall increase or decrease in code size, providing insight into whether the patch involves extensive restructuring or more targeted, minimal modifications.
\end{itemize}

\textit{Statistical Analysis:} We apply the \textit{Mann-Whitney U Test}, a non-parametric method suitable for comparing independent groups with non-normal distributions~\cite{wang2023delving}, to assess the significance of metric differences and identify factors affecting issue resolution.

\subsection{\textbf{Issue Analysis Results: Resolved vs. Unresolved}}

Table~\ref{tab:github_issue_analysis} presents the results of comparing 330 resolved and 170 unresolved GitHub issues, focusing on problem statements, associated source code files, and gold patch solutions.
\vspace{-1.2em}
\begin{table}[htbp]
    \centering
    \caption{GitHub Issue Analysis: Resolved vs. Unresolved Issues}
    \resizebox{\columnwidth}{!}{%
    \begin{tabular}{lrrrr}
        \toprule
        \textbf{Metric} & \textbf{Resolved Mean} & \textbf{Unresolved Mean} & \textbf{Mann-Whitney U Test p-value} & \textbf{Significance} \\
        \midrule
        \multicolumn{5}{c}{\textbf{1. Problem Statements}} \\
        \midrule
        Flesch\_Reading\_Ease & 35.63 & 38.91 & 0.2425 & \textcolor{red}{\xmark} \\
        Flesch\_Kincaid\_Grade & 11.40 & 10.81 & 0.2219 & \textcolor{red}{\xmark} \\
        Sentence\_Count & 27.46 & 37.22 & 0.0175 & \textcolor{darkgreen}{\checkmark} \\
        Word\_Count & 178.72 & 209.86 & 0.0015 & \textcolor{darkgreen}{\checkmark} \\
        \midrule
        Contains\_Code\_Snippets & 0.45 & 0.48 & 0.5130 & \textcolor{red}{\xmark} \\
        Number\_of\_Code\_Blocks & 1.08 & 1.18 & 0.5295 & \textcolor{red}{\xmark} \\
        Lines\_of\_Code & 14.44 & 18.56 & 0.47 & \textcolor{red}{\xmark} \\
        Code\_to\_Text\_Ratio & 0.24 & 0.24 & 0.7960 & \textcolor{red}{\xmark} \\
        \midrule
        \multicolumn{5}{c}{\textbf{2. Associated Source Code Files (Gold Patch Based)}} \\
        \midrule
        Code\_Files\_Count & 1.09 & 1.53 & 3.53e-11 & \textcolor{darkgreen}{\checkmark} \\
        Lines\_of\_Code & 703.13 & 1087.38 & 0.0062 & \textcolor{darkgreen}{\checkmark} \\
        Code\_\_Cyclomatic\_Complexity & 192.25 & 303.12 & 0.0067 & \textcolor{darkgreen}{\checkmark} \\
        \midrule
        \multicolumn{5}{c}{\textbf{3. Gold Patch Solutions}} \\
        \midrule
        Total\_Lines\_Change & 9.29 & 24.12 & 1.67e-09 & \textcolor{darkgreen}{\checkmark} \\
        Net\_Code\_Size\_Change & 3.22 & 10.08 & 9.18e-06 & \textcolor{darkgreen}{\checkmark} \\
        \bottomrule
    \end{tabular}%
    }
    \label{tab:github_issue_analysis}
\end{table}
\vspace{-0.9em}

\textit{Problem Statements:} Resolved issues had shorter sentences and fewer words, suggesting agents perform better with concise problem descriptions. Readability metrics like \textit{Flesch Reading Ease} and \textit{Flesch-Kincaid Grade}, as well as the inclusion of code snippets, showed no significant differences, indicating that neither readability nor the presence of code strongly impacts issue resolution.

\textit{Associated Source Code Files:} Resolved issues involved fewer and less complex source code files, with a significantly lower number of modified lines of code and lower code cyclomatic complexity, suggesting that agents perform better when the task involves smaller and simpler codebases.
  
\textit{Gold Patch Solutions:} 
Resolved issues had a significantly smaller total lines change (mean = 9.29) compared to unresolved issues (mean = 24.12). Similarly, the net code size change was lower for resolved issues than for unresolved ones. This indirect evidence suggests that unresolved issues require more extensive modifications and are more challenging to resolve.

\vspace{-0.5em}
\begin{tcolorbox}[colback=yellow!10!white, colframe=yellow!40!black, title=Finding 9] Resolved GitHub issues generally involved fewer files, smaller codebases, and more modest code changes, indicating that agents are more effective at handling simpler tasks. However, the non-significant differences in metrics like readability and the presence of code snippets suggest these are not the main factors influencing agent performance. These findings imply that breaking down complex issues into smaller, more manageable tasks could improve agent performance and enhance their overall effectiveness. \end{tcolorbox}
\vspace{-0.5em}

\label{sec:discuss}
\section{Discussion}
\subsection{Suggestions}
\subsubsection{\textbf{Suggestions for SWE-Bench Verified Maintainers}}
Enhance the evaluation framework by expanding test coverage to detect potential side effects from agent-generated patches that diverge from gold patches, even if unit tests pass. Incorporate code quality metrics such as complexity, duplication, and code smells into the assessment criteria to ensure agents produce maintainable code~\cite{liu2024your,velasco2024beyond}. 

\subsubsection{\textbf{Suggestions for AI Agent Developers}} Enhance agents' ability to handle complex tasks by breaking down issues into manageable sub-tasks. While improving functional correctness, focus on the non-functional aspects of the generated solutions, such as avoiding over-modifications, improving maintainability, reducing code complexity and duplication, and ensuring no new bugs or vulnerabilities are introduced. Consider integrating additional safeguards, like \textit{Code Shield}, to promote secure software development~\cite{he2023large,kavian2024llm}.


\subsubsection{\textbf{Suggestions for Users of AI Agents}}
Utilize multiple agents to leverage their varied strengths, increasing the chances of successful issue resolution. Carefully review agent-generated patches for over-modifications and potential unintended effects, even if they pass unit tests. For complex issues, consider decomposing them into smaller, manageable tasks to align with agents' current capabilities.

\subsection{Threats to Validity}

\subsubsection{\textbf{Threats to Internal Validity}} Our study may be affected by data collection bias and measurement reliability. Since we relied on data from the SWE-Bench Verified dataset and agent-generated patches, any errors in data extraction or processing could influence the results. To mitigate this, we automated data collection and performed manual checks for accuracy. Additionally, using \textit{SonarQube} for code quality analysis could introduce measurement errors. We addressed this by using the widely accepted community version of \textit{SonarQube} and ensuring consistent analysis conditions.

\subsubsection{\textbf{Threats to External Validity}} Our findings may have limited generalizability, as the study focuses on 500 GitHub issues, which may not represent other programming languages or project types. However, SWE-Bench Verified, with 12 diverse and widely-used Python repositories, strengthens the relevance of our results. As future benchmarks expand to more languages and project scenarios, we plan to extend our study accordingly. While software development agents evolve rapidly, the data from the public leaderboard reflects the most recent rankings at the time of data collection, ensuring the timeliness of our analysis.

\subsubsection{\textbf{Threats to Construct Validity}} 
The validity of our metrics and comparisons may pose a threat. Metrics like code smells, cyclomatic complexity, and our statistical tests may not capture all aspects of code quality or agent performance. To mitigate this, we used well-established metrics and multiple measures for a comprehensive assessment. Comparing agent patches to gold patches assumes the gold patches are optimal, which may not always be the case. We addressed this by also evaluating the impact of agent patches on code quality, acknowledging that alternative solutions can be acceptable if they maintain or improve quality.

\section{Related Work}
Recent studies have explored the security and quality of code generated by large language models (LLMs) like GitHub Copilot and ChatGPT. Pearce et al.~\cite{pearce2022asleep} found that approximately 40\% of Copilot-generated code was vulnerable to CWE Top 25 weaknesses, while a replication by Majdinasab et al.~\cite{majdinasab2024assessing} reduced this to 27.25\%, highlighting ongoing security concerns. Asare et al.~\cite{asare2023github} and Hamer et al.~\cite{hamer2024just} compared LLM-generated code with human-written code and StackOverflow snippets, noting that while LLMs can introduce vulnerabilities, they sometimes perform comparably or better than human developers. Nguyen et al.~\cite{nguyen2022empirical} analyzed Copilot's code suggestions using LeetCode problems, revealing variations across languages, along with issues like high complexity and reliance on undefined methods. Similarly, Liu et al.~\cite{liu2024no} and Liu et al.~\cite{liu2024refining} assessed ChatGPT's performance on algorithmic tasks and found issues in code correctness and maintainability. Rabbi et al.~\cite{rabbi2024ai} and Siddiq et al.~\cite{siddiq2024quality} further emphasized the challenges in using ChatGPT-generated code, identifying limitations in quality and maintainability.

Our research offers two notable contributions that differentiate it from related work. First, we evaluate the quality of code produced by \textit{software development agents}—like Amazon-Q Developer Agent and AppMap Navie + GPT 4o—that enhance LLM capabilities through agentic workflows beyond 
standalone 
or base LLMs. Second, unlike prior work focusing on simplified scenarios like isolated algorithmic challenges or vulnerability-prone prompts, we assess code quality with \textit{real-world GitHub issues}, which involve complex codebases and require modifications across multiple files. This provides a more realistic evaluation of agent-generated code, bridging a critical gap in the literature.

\section{Conclusion}
This study 
analyzed
4,892 patches generated by 10 software development agents on 500 real-world GitHub issues from SWE-Bench Verified, focusing on their impact on code quality. No single agent dominated, with 170 issues unresolved, highlighting areas for improvement. 
Even for patches that passed unit tests and resolved issues, their divergence from ``gold patches'' revealed risks not captured by current tests. While some agents like \textit{Gru} demonstrated more balanced modifications, and the others like \textit{HoneyComb} over-modified the code, impacting maintainability. Most agents maintained code reliability and security, avoiding new bugs or vulnerabilities, and performed comparably to human patches in reducing code smells and duplication. However, some agents need improvement in minimizing code complexity and duplication. Lastly, agents were more successful with simpler tasks, suggesting that breaking down complex issues could enhance their effectiveness.

Future work should focus on improving agents' ability to handle more complex scenarios, as well as expanding the benchmarks to include vulnerability-prone issues for a deeper evaluation of agent performance in secure software development.

\vspace{-0.5em}
\begin{tcolorbox}[colback=yellow!10!white, colframe=yellow!40!black, boxrule=0.4pt, left=3pt, right=3pt, top=3pt, bottom=3pt]
\small
\textbf{Replication Package}: To facilitate further research and enable reproducibility, we provide datasets and scripts at: \url{https://osf.io/5urgc/?view_only=210932a785204432b86d857c089e25dd}.
\end{tcolorbox}
\vspace{-1em}

\section*{Acknowledgments}
This research is supported by the Ministry of Education, Singapore under its Academic Research Fund Tier 3 (Award ID: MOET32020-0004). Any opinions, findings and conclusions or recommendations expressed in this material are those of the author(s) and do not reflect the views of the Ministry of Education, Singapore.

%

\onecolumn
\begin{multicols}{2}
\bibliographystyle{IEEEtran}
\bibliography{main}
\end{multicols}

\end{document}